\documentclass[11pt,a4paper]{article}
\usepackage{jheppub}

\usepackage{epsfig}
\usepackage{amssymb}

\def\ba{\begin{eqnarray}}
\def\ea{\end{eqnarray}}

\def\fun#1#2{\lower3.6pt\vbox{\baselineskip0pt\lineskip.9pt
  \ialign{$\mathsurround=0pt#1\hfil##\hfil$\crcr#2\crcr\sim\crcr}}}
\def\order#1{{\mathcal O}\left(#1\right)}
\newcommand{\vecc}[1]{\mbox{\boldmath $#1$}}
\def\gsim{\mathrel{\raise.3ex\hbox{$>$\kern-.75em\lower1ex\hbox{$\sim$}}}}
\def\lsim{\mathrel{\raise.3ex\hbox{$<$\kern-.75em\lower1ex\hbox{$\sim$}}}}

\title{On relativization of the Sommerfeld-Gamow-Sakharov factor}

\author[a,b]{Andrej B. Arbuzov}
\emailAdd{arbuzov@theor.jinr.ru}
\author[b]{Tatiana V. Kopylova}

\affiliation[a]{BLTP, JINR, Joliot-Curie 6, Dubna, 141980 Russia}
\affiliation[b]{Department of Higher Mathematics,
University of Dubna,  141980 Dubna, Russia}

\abstract{
The Sommerfeld-Gamow-Sakharov factor is considered for the general case
of arbitrary masses and energies. It is shown that the scalar triangular
one-loop diagram gives the Coulomb singularity in radiative corrections
at the threshold. The singular part of the correction is factorized at
the complete Born cross section regardless of its partial wave decomposition.
Different approaches to generalize the factor are discussed. 
}

\keywords{
Coulomb singularity, final state interactions, rescattering}

\begin{document}

\maketitle

\section{Introduction}
\label{Intro}

It is well known that the electromagnetic interaction between charged particles
in the final state can considerably affect the observable reaction rate.
For example, the cross section of electron-positron annihilation into muons
becomes different from zero at the threshold due to the final state interactions.
Another observable effect is the difference in energy behavior at the threshold
of the annihilation channels with production of charged and neutral mesons,
see paper~\cite{Voloshin:2003gm} and references therein.
In the case of strong interactions in the final state, e.g. in
the processes with creation of a heavy quark pair, a similar nonperturbative
enhancement factor appears~\cite{Brodsky:1995ds}.
It was shown~\cite{Anchishkin:1994sa} that 
interplay of both QED and QCD final state interactions can be also important.
The effects of the Coulomb singularity in production of stop and gluino pairs
close to their thresholds were discussed in Ref.~\cite{Bigi:1991mi}. 

If the relative velocity of the charged particles is small $(v\ll 1)$\footnote{We use
natural units $c=\hbar=1$.}, then
the effect of multiple photon exchange between them becomes significant. 
This fact has been discussed in the literature for a long time. It was shown
already in the textbook by A.~Sommerfeld~\cite{Sommerfeld:1921book} that the
correction due to re-scattering of charged particles in the final state is proportional
to the wave function at the origin squared, $|\Psi(0)|^2$, 
see also book~\cite{Schwinger:book}. 
So that the scattering (or annihilation) channel acquires some features 
of the corresponding bound state.
G.~Gamow has shown~\cite{Gamow:1928zz} that the same factor is relevant
for the description of the Coulomb barrier in nuclear interactions. 
Using the non-relativistic Schr\"odinger equation, 
A.~Sakharov derived this factor for the case of charged pair 
production~\cite{Sakharov:1948yq} in the form 
\ba \label{factor}
T = \frac{\eta}{1-e^{-\eta}},\qquad \eta =\frac{2\pi\alpha}{v}\, ,
\ea
where $\alpha\approx 1/137$ is the fine structure constant and
$v$ is the (non-relativistic) relative velocity of the particles in the created pair,
\ba
v = \left|\frac{\vec{p}_1}{m_1}-\frac{\vec{p}_2}{m_2}\right|.
\ea
Here $\vec{p}_{1,2}$ and $m_{1,2}$ are the momenta and masses of the particles.

\begin{figure}
\includegraphics[width=9.2cm,height=6cm,angle=0]{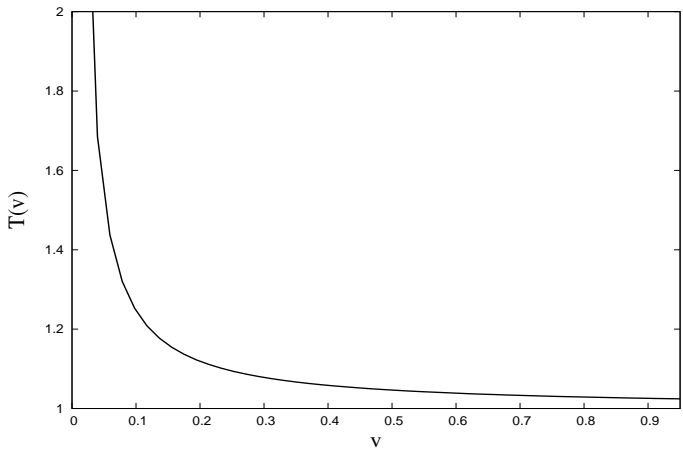}
\caption{SGS factor in QED vs. the relative velocity.
\label{fig:factor_v}}
\end{figure}

The behavior of the factor is shown in Fig.~\ref{fig:factor_v}

The question about relativization of the Sommerfeld-Gamow-Sakharov (SGS) factor and about
some other ways of its generalization, e.g. for non-equal masses
and $P$-waves, is under discussion in the literature for a long time, 
see papers~\cite{baier:1969ccc,Fadin:1988fn,Fadin:1990wx,Bardin:1993mc,Arbuzov:1993qc,Yoon:1999jd,Yoon:2004sr,Solovtsova:2009zq} 
and references therein.
In ref.~\cite{Castellani:2005bf} the possibility to generalize the factor by 
inclusion of strong and weak interactions was studied.

\section{SGS factor}
\label{SGS_fact}

Let us first discuss the general features of the SGS factor.
It is useful to consider the limit of a small coupling 
constant\footnote{Obviously, this expansion can not be used if $v\leq 2\pi\alpha$.}
\ba \label{sgs_expand}
\lim_{\eta\to 0} T = 1 + \frac{\eta}{2} + \frac{\eta^2}{12} + \order{\eta^3}
= 1 + \frac{\pi\alpha}{v} + \frac{\pi^2\alpha^2}{6v^2} + \order{(\alpha/v)^3}.
\ea
In this way we get terms which can be related to the ones arising in
a perturbative calculation.

Formula~(\ref{factor}) can be easily adapted for the case of 
arbitrary charges $Q_1$ and $Q_2$ by taking $\eta = -Q_1Q_2 \cdot 2\pi\alpha/v$. 
The fact that for the repulsion case $(Q_1Q_2>0)$ there is no Coulomb singularity
provides an asymmetry in contributions of different quarks pairs taken from the
final state hadrons. This allows to discuss in refs.~\cite{Baldini:2007qg,Baldini:2008nk} 
the possibility the have a threshold enhancement factor even for the neutral 
baryon $(\Lambda\bar\Lambda)$ production case.

Let us consider the case of the final state 
interaction\footnote{A very similar picture takes place 
for the case of initial state interactions.}
of two charged particles produced close to the threshold, e.g.
in electron-positron annihilation
\ba
&& e^-(k_1)\ +\ e^+(k_2)\ \to\ a^-(p_1)\ +\ a^+(p_2),
\\
&& 
s = (k_1+k_2)^2 = (p_1+p_2)^2 \gsim (m_1+m_2)^2, 
\ea
where $a^{\pm}$ can be scalar, spinor, or vector particles. 
The Born-level cross section $\sigma^{\mathrm{Born}}$ of this process 
depends on the type of integration and spin. 
But in any case in the center-of-mass system, 
it is proportional to the first power of factor $\beta_{1,2}$ which
comes from the phase space volume 
and vanishes at the threshold $s\to(m_1+m_2)^2$,
\ba
&& \beta_{1,2} = \frac{2p}{p_1^0+p_2^0}, 
\qquad  p \equiv |\vecc{p}_1|=|\vecc{p}_2|=\frac{\sqrt{\Lambda(s,m_1^2,m_2^2)}}{2\sqrt{s}}, 
\nonumber \\
&& p_1^0+p_2^0= 2\sqrt{s}, \qquad  
\Lambda(x,y,z) = x^2+y^2+z^2-2xy-2xz-2yz.
\ea
For the case of equal masses this factor takes the usual
form of the relativistic velocity $\beta=\sqrt{1-m^2/(p^0)^2}$ of the final state 
particles. 

\begin{figure}
\begin{center}
\includegraphics[width=5.0cm,angle=0]{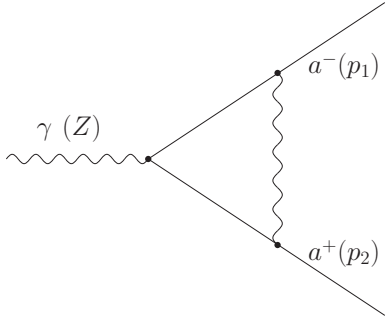}
\end{center}
\caption{Feynman diagram for one-loop virtual correction in the final state.
\label{fig:1}}
\end{figure}

In the one-loop approximation the QED radiative correction to the 
annihilation cross section gets contributions from virtual and real photon
emission:
\ba
\sigma^{\mathrm{1-loop}} = \sigma^{\mathrm{Born}}\biggl(1 + \delta^{\mathrm{Virt}} 
+ \delta^{\mathrm{Real}}\biggr).
\ea
The last term in the parentheses is proportional to $\beta^2$, it is strongly suppressed
at the threshold. Explicit expressions for different contributions 
with the exact dependence on the final state particle mass (for the equal mass case)
can be found e.g. in ref.~\cite{Arbuzov:1991pr} for fermions and 
in ref.~\cite{Arbuzov:1997je} for scalars. The final state one-loop virtual correction
(in the on-mass-shell renormalization scheme) is described by the triangle diagram
shown in Fig.~\ref{fig:1}. There are three types of integrals over the loop momentum:
the scalar, the vector and the tensor ones:
\ba \nonumber
\left\{I_S,\, I_V^{\mu},\, I_T^{\mu\nu}\right\} = \int\frac{d^4k}{i\pi^2}\;
\frac{\{ 1,\; k^\mu,\; k^{\mu\nu}\}}
{((p_1+k)^2-m_1^2+i\varepsilon)((p_2-k)^2-m_2^2+i\varepsilon)(k^2+i\varepsilon)}\, .
\ea
The tensor one contains an ultraviolet divergence,
which has to be removed by the standard renormalization procedure. 
The vector integral $I_V^{\mu}$ is finite.
The scalar integral $I_S$  contains an infrared divergence, which cancels out in the sum
with the contribution of the real final state radiation.

Direct calculations show that the contributions of the vector and tensor integrals
are suppressed by at least the first power of the final state particle velocities. 
While the scalar integral is proportional to $1/\sqrt{\Lambda(s,m_1^2,m_2^2)}$
and reveals at the threshold the well known Coulomb singularity. 
The coefficients before the integrals depend in general on the type of the particles,
but the factor standing at the scalar integral is {\em universal}, it is the same 
for scalar, spinor and vector final state particles. The contributions of 
the one-loop scalar integral to the cross section can be presented in the form
\ba \label{1-loop}
\delta\sigma^{\mathrm{1-loop}}_S = \sigma^{\mathrm{Born}} 
\frac{\alpha}{\pi}\, Q_1Q_2 (s-m_1^2-m_2^2) I_s, \\ \nonumber 
I_s \equiv C_0(m_1^2,m_2^2,s,m_1^2,m_\gamma^2,m_2^2),
\ea
where the notation of the LoopTools package~\cite{Hahn:1998yk} 
for the Passarino-Veltman functions is used.
The explicit form of this integral can be found for example in ref.~\cite{Bardin:1993mc}. 
The infrared divergence of this integral can be regularized 
by a fictitious photon mass $m_\gamma$ or with the help of any other 
regularization scheme.

At the threshold,  the integral takes the simple form
\ba \label{C0_exp}
\lim_{s\to(m_1+m_2)^2}I_s = \frac{1}{\sqrt{\Lambda(s,m_1^2,m_2^2)}}\left[
- \pi^2 + \order{\sqrt{\frac{s-(m_1+m_2)^2}{s}}} \right].
\ea
Comparison of the order $\alpha$ term in Eq.~(\ref{sgs_expand}) with the one 
obtained above gave us a hint to make the choice of SGS factor generalization.
Namely, we see that the one-loop calculation is consistent with the substitution
of the non-relativistic relative velocity by its relativistic version
\ba \label{v_R}
v_{\mathrm{rel}} = \frac{\sqrt{\Lambda(s,m_1^2,m_2^2)}}{s-m_1^2-m_2^2} 
= \frac{\sqrt{[s-(m_1+m_2)^2][s-(m_1-m_2)^2]}}{s-m_1^2-m_2^2}\, .
\ea
We would like to underline that $v_{\mathrm{rel}}$ is exactly the relativistic
sum of the velocities of our particles. This quantity is a relativistic invariant. 
For $s\gg m_{1,2}^2$ in the ultra-relativistic limit $v_{\mathrm{rel}}\to 1$.

In the limiting case when one of the masses is heavy and the other is light,
the relative velocity coincides with the one of the light particle (in the rest
reference frame of the heavy particle). In this case the relativized
SGS factor emerges from the relativistic one-particle Dirac (or Klein--Gordon--Fock) equation
in a central field.

\section{Discussion}

The SGS factor is a part of radiative corrections which are used 
in the analysis of modern experimental data on various processes. 
To avoid a double counting we should {\em match} the factor with
other higher order QED contributions. We suggest to perform the matching
in the following way:
\ba \label{eq:match}
\sigma^{\mathrm{Corr.}} &&= \sigma^{\mathrm{Born}}\biggl(T(v)
 - \frac{\pi\alpha}{v}
 - \frac{\pi^2\alpha^2}{3v^2} - \ldots \biggr)
\nonumber \\ \nonumber 
&&  + \Delta\sigma^{1-loop}
 +\Delta\sigma^{2-loop} + \ldots
\ea
Here $\Delta\sigma^{n-loop}$ is the $n$-th loop perturbative QED contribution to
the observed (corrected) cross section $\sigma^{\mathrm{Corr.}}$. The Born level
cross section $\sigma^{\mathrm{Born}}$ can be taken in an {\em improved} approximation:
it may include some higher order effects not related (to interactions in the final state), 
{\it e.g.} the vacuum polarization and the initial state radiation.
So we subtract from the factor the first orders of its perturbative expansion,
which are then restored by adding the explicit perturbative results (in the same
orders). 

\begin{figure}
\begin{minipage}[t]{0.49\textwidth}
\includegraphics[width=7.2cm,angle=0]{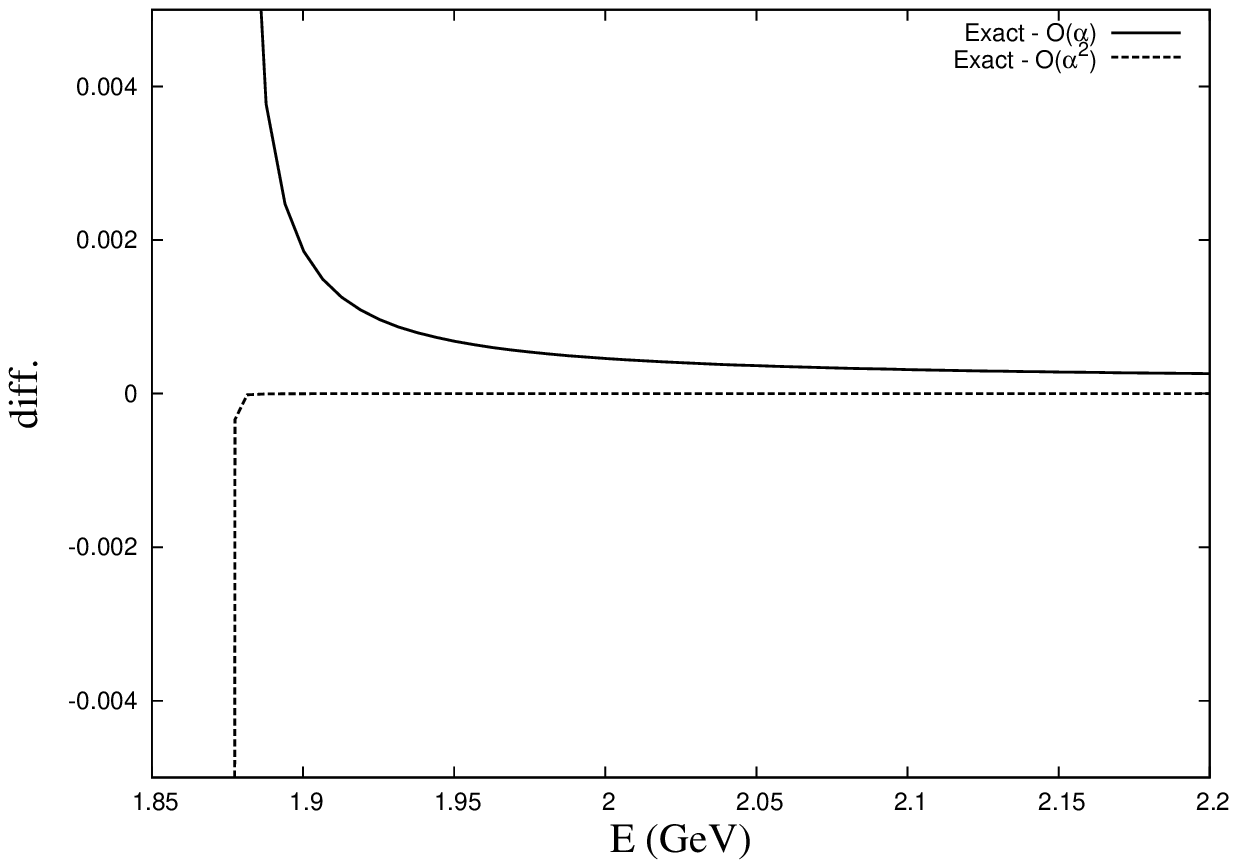}
\caption{Difference between the resummed SGS factor and its perturbative
expansion in different approximations.
\label{fig:diff_exp}}
\end{minipage}
  \hfill
\begin{minipage}[t]{0.49\textwidth}
\includegraphics[width=7.2cm,angle=0]{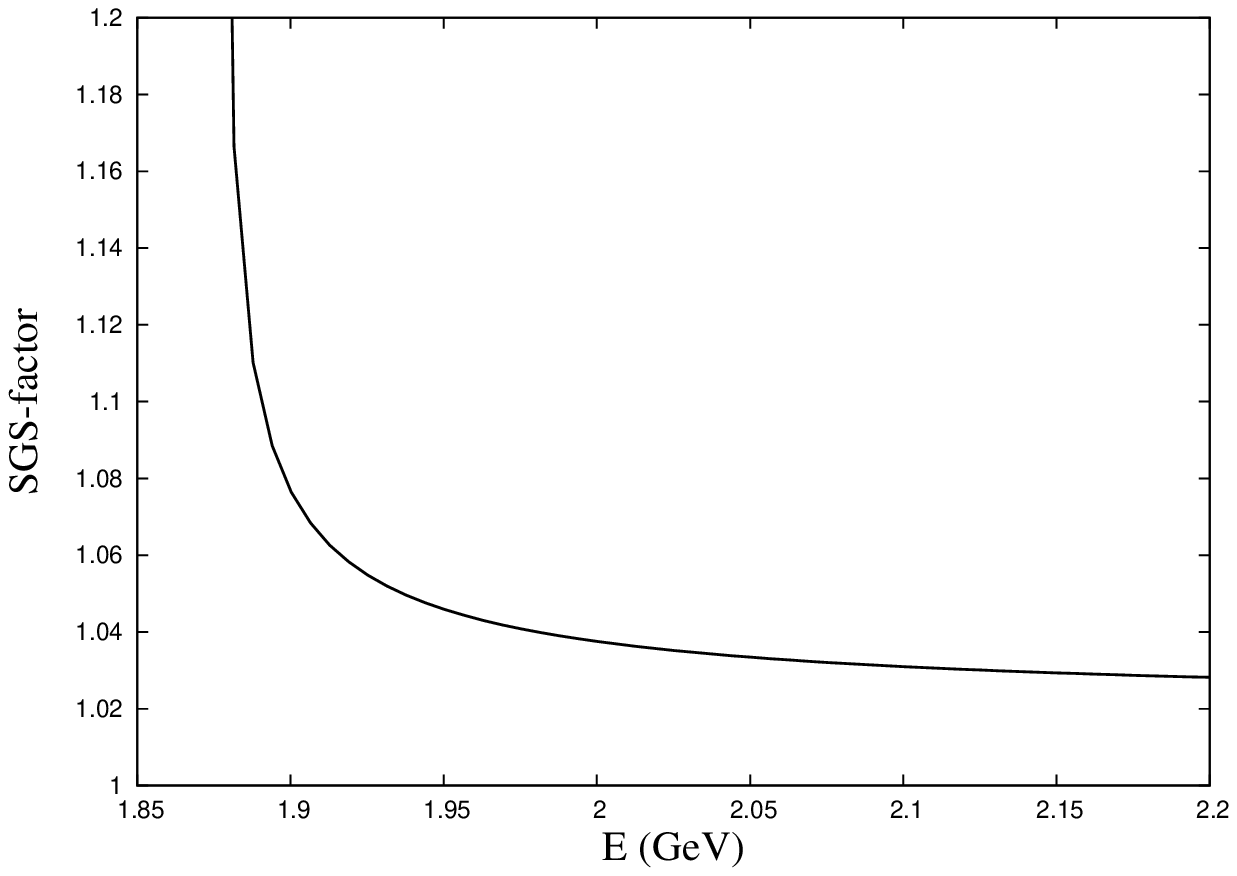}
\caption{SGS factor vs. the center-of-mass energy for $e^+e^-\to p\bar{p}$.
\label{fig:factor}}
\end{minipage}
\end{figure}

Fig.~\ref{fig:diff_exp} shows the difference between the complete SGS factor~(\ref{factor}) and its
perturbative expansion~(\ref{sgs_expand}) for the process $e^+e^-\to p\bar{p}$ as a function of the
center-of-mass energy. One can see that the difference is steeply rising up (or going down) in the
region close to the threshold where the perturbative expansion breaks down. Nevertheless, soon above
the threshold the difference becomes small especially for the case of the $\order{\alpha^2}$ 
approximation.

Let us compare our version of the generalized SGS factor with the other ones
known in the literature. 
First of all, we note that the expression of the factor obtained here by
extrapolation of the one-loop result coincides with the one derived
in ref.~\cite{Arbuzov:1993qc} for the case of scalar particles with 
the help of a relativistic quasi-potential equation. Analyzing another
relativistic quasipotential equation suggested by I.~Todorov
in ref.~\cite{Todorov:1970gr} (see eq.~(4.1) there) also gives the same 
value of the wave function in the origin and thus the same version of the
relativized SGS factor.

In ref.~\cite{baier:1969ccc}  resummation of ladder diagrams in the final
state interactions of equal-mass fermions was performed. It was explicitly demonstrated
that the form~(\ref{factor}) of the SGS factor is reproduced. But instead of the 
non-relativistic relative velocity quantity $2\beta$ emerged (the same as in the 
{\it ad hoc} relativization procedure). To our mind, the reason for this is as follows.
Keeping only the ladder diagram contribution without crossed photon lines corresponds
to the pure Coulomb photon exchange, while in the relativistic case its contribution
has the same order as the one due to transverse photons. It has been demonstrated 
in ref.~\cite{Arbuzov:1993qc} within a quasi-potential relativistic equation approach that
keeping only the Coulomb interaction in the potential leads to $v=2\beta$ while
adding the transverse photon exchange restores the complete value 
of the relativistic relative velocity.  
Note also that the {\it ad hoc} relativized version of the SGS factor $(v=2\beta)$
has a wrong ultra-relativistic limit $v\to 2$. Moreover, this version of the factor
it is not relativistic invariant.

Our version of the SGS factor is also supported in ref.~\cite{Hoang:1997sj}, where
the explicit analytical results for final state QED corrections to production
of spinor particle with equal masses were considered. It was shown that the
relativistic relative velocity 
\ba
v=\frac{2\beta}{1+\beta^2}
\ea
naturally appears in the case considered.

An original version of the relativized SGS factor for the case of arbitrary masses
was derived in ref.~\cite{Solovtsova:2009zq} with the help of a relativistic 
two-body equation. This version of the factor satisfies the main condition: the
non-relativistic expression is reproduced at the threshold. But the ultra-relativistic
limit and the heavy--light mass $(m_2\gg m_1)$ one for $v$ are missed.

Authors of ref.~\cite{Ferroli:2010bi} claimed that the experimental data on the
process $e^+e^-\to p\bar{p}$ favor application of the SGS factor in the threshold region
without its denominator 
\ba \label{e_factor}
T\bigg|_{v\ll 1} \approx {\mathcal E}= \frac{\pi\alpha}{\beta}.
\ea 
In this case the multiplier
\ba \label{r_factor}
{\mathcal R}=\frac{1}{1-e^{-\pi\alpha/\beta}}
\ea 
called there as the {\em resummation factor} is dropped. 
We would like to note, that the introduction of the resummation 
factor looks rather artificial in the view of the perturbative
expansion~(\ref{sgs_expand}), i.e. the enhancement factor~(\ref{e_factor}) 
contains itself a certain nonperturbative (resummed) contribution.  
Moreover, neither the known one-loop QED corrections, nor the advocated above
form of the relativized SGS factor were applied there.

In ref.~\cite{Smith:1993vp} higher order effects due to the fine structure constant 
running were taken into account (see eq.(18) there) in the form of an additional factor. 
It is clear that this effect becomes numerically important only for ultra-relativistic 
relative velocities. The additional factor derived in~\cite{Smith:1993vp} can be also
applied for our version of the SGS factor. Another possibility is to take into
account the running of $\alpha$ in the perturbative part of the
matching formula~(\ref{eq:match}).  

In ref.~\cite{Bardin:1993mc} application of the SGS factor to production of
unstable charged particles (a $W^\pm$ pair) was considered. The authors also
evaluated the one-loop triangular diagram. The factor $(s-m_1^2-m_2^2)$ 
before the one-loop integral, see (\ref{1-loop}), was approximated there to be equal $s/2$.
For this reason, their expression for the {\em relativized} relative velocity
(standing in the SGS factor)
has the correct non-relativistic limit but does not satisfy the ultra-relativistic
and heavy-light mass limits.

Papers~\cite{Fadin:1993kg,Fadin:1994pm,Fadin:1995fp} also discuss production of $W^+W^-$
boson pair near the threshold taking into account the width of $W$ bosons and some higher
order corrections. The relative velocity in the SGS factor was treated there in a  
non-relativistic manner: $v=v^++v^-$, where $v^\pm$ was either relativistic 
or non-relativistic velocity of $W^\pm$ in the c.m.s. We would like to underline that
at the threshold such an approximation is very solid.

\section{Conclusions}

As discussed above there are several different approaches
to generalize the Sommerfeld-Gamow-Sakharov factor.
Most of them are equivalent from the practical point of view,
since they differ by terms that vanish in the limit $v\to0$.
On the other hand, such terms are not universal (they depend on
the choice of the process) and can not be re-summed in a unique way. 

We demonstrated that there is a certain part of 
the final state QED correction which does not depend on spin
of the interacting particles and appear in the order-by-order
perturbative calculations exactly in the form of the non-relativistic
factor expansion. The corresponding recipe of the SGS factor relativization
consists just in the substitution of the non-relativistic relative
velocity of the two particles by the relativistic one. This choice  
could have been suggested from the beginning, but actually we got
it here by looking at one-loop perturbative radiative corrections. 
It is worth to underline that exactly the same relativized SGS factor 
was received in refs.~\cite{Todorov:1970gr,Arbuzov:1993qc} 
with the help of relativistic quasi-potential equations. 

The widely used version of the relativized SGS factor where 
the non-relativistic relative velocity is substituted by $2\beta$
(twice the velocity of a particle in the center-of mass frame)
was criticized. In fact this version of the factor is not 
relativistic invariant and has a wrong ultra-relativistic limit.

As concerning phenomenological applications, we claim that any
choice of the SGS factor which has the correct non-relativistic 
limit can be used. One should just take care on removing possible
double counting if other ({\it e.g.} complete one-loop) radiative corrections
are taken into account. The uncertainty due to the choice of the concrete 
SGS factor will lie then in uncontrolled terms of higher orders in $\alpha$,
which are not singular in the limit $v\to0$. 

For the case of non-equal masses we suggested to verify the 
heavy--light mass limit $m_1\gg m_2$, where one can use for a cross check
not only the solution of the non-relativistic Schr\"odinger equation
but also the one of the relativistic Dirac equation. Our choice of the SGS factor
satisfies this condition by construction.

It is worth noting that the SGS factor derived here is applicable for
all partial waves which can appear in the process $\gamma^*\to ab$. In particular,
it is factorized before the whole Born cross section of $e^+e^-\to p\bar{p}$
(before both the $S$ and $D$ wave contributions to it). For the case of pseudoscalar
meson production, {\it e.g.} for $e^+e^-\to\pi^+\pi^-$, the same factor stands for $p$ 
wave. Of course, the $P$ and $D$ waves Born-level contributions are proportional to higher
powers of $\beta$, so there is no shift from $0$ of the cross section at the threshold 
due to the Coulomb enhancement, but the radiative corrections themselves 
are large in that region.

The version of the SGS factor advocated here is implemented into Monte Carlo
code {\tt MCGPJ}~\cite{Arbuzov:2005pt} with matching to the complete first order
corrections.
Obviously, taking into account the resummation factor in the proper
approximation would be important for the analysis of new 
precise data on production of particles near threshold
are coming from experiments at VEPP2000 (Novosibirsk),
BEPCII (Beijing), and other machines.

\acknowledgments{
We are grateful to D.~Bardin and E.~Kuraev for useful discussions. 
One of us (A.A.) was supported by the RFBR grant 10-02-01030-a.}

\end{document}